\documentclass[sigconf]{acmart}
\usepackage{amsmath}
\usepackage{multirow}
\usepackage{algorithm}
\usepackage{algpseudocode}
\usepackage{tablefootnote}
\usepackage{booktabs}
\usepackage{lipsum}
\usepackage{tabularray}
\usepackage[normalem]{ulem}
\useunder{\uline}{\ul}{}

\AtBeginDocument{%
  \providecommand\BibTeX{{%
    \normalfont B\kern-0.5em{\scshape i\kern-0.25em b}\kern-0.8em\TeX}}}

\acmConference[Conference acronym 'XX]{Make sure to enter the correct
  conference title from your rights confirmation emai}{June 03--05,
  2018}{Woodstock, NY}
\begin{document}

\title{Beyond Relevance: Improving User Engagement by Personalization for Short-Video Search}

% \author{Wentian Bao, Hu Liu, Kai Zheng, \\ Shunyu Zhang,  Tong Guo, Chao Zhang, Yun En Yu, Wenwu Ou, Yang Song}
% \affiliation{%
%   \institution{Kuaishou Technology}
%   \city{Beijing}
% \email{{zhengkai,zhangshunyu,guotong03,zhangchao,wangyong07,yangsong}@kuaishou.com}
% \email{hooglecrystal@126.com,{wentian0262,ouwenwu}@gmail.com}
% % \authornote{corresonding author.}

\author{Wentian Bao}
\affiliation{%
  \institution{unaffiliated}
  \country{China}}
\email{wentian0262@gmail.com}

\author{Hu Liu}
\affiliation{%
  \institution{Kuaishou Technology}
  \city{Beijing}
  \country{China}}
\email{hooglecrystal@126.com}

\author{Kai Zheng}
\affiliation{%
  \institution{Kuaishou Technology}
  \city{Beijing}
  \country{China}}
\email{zhengkai@kuaishou.com}

\author{Chao Zhang}
\affiliation{%
  \institution{Kuaishou Technology}
  \city{Beijing}
  \country{China}}
\email{zhangchao@kuaishou.com}

\author{Shunyu Zhang}
\affiliation{%
  \institution{Kuaishou Technology}
  \city{Beijing}
  \country{China}}
\email{zhangshunyu@kuaishou.com}

\author{Enyun Yu}
\affiliation{%
  \institution{unaffiliated}
  \country{China}}
\email{yuenyun@126.com}

\author{Wenwu Ou}
\affiliation{%
  \institution{unaffiliated}
  \country{China}}
\email{ouwenwu@gmail.com}

\author{Yang Song}
\affiliation{%
  \institution{Kuaishou Technology}
  \city{Beijing}
  \country{China}}
\email{yangsong@kuaishou.com}

\renewcommand{\shortauthors}{Wentian, et al.}


\begin{abstract}
Personalized search has been extensively studied in various applications, including web search, e-commerce, social networks, etc. With the soaring popularity of short-video platforms, exemplified by TikTok and Kuaishou, the question arises: can personalization elevate the realm of short-video search, and if so, which techniques hold the key?

In this work, we introduce \textbf{$\text{PR}^2$}, a novel and comprehensive solution for personalizing short-video search, where $\text{PR}^2$ stands for the \textbf{P}ersonalized \textbf{R}etrieval and \textbf{R}anking augmented search system. Specifically, $\text{PR}^2$ leverages query-relevant collaborative filtering and personalized dense retrieval to extract relevant and individually tailored content from a large-scale video corpus. Furthermore, it utilizes the \textbf{QIN} (\textbf{Q}uery-Dominate User \textbf{I}nterest \textbf{N}etwork) ranking model, to effectively harness user long-term preferences and real-time behaviors, and efficiently learn from user various implicit feedback through a multi-task learning framework. By deploying the $\text{PR}^2$ in production system, we have achieved the most remarkable user engagement improvements in recent years: a 10.2\% increase in CTR@10, a notable 20\% surge in video watch time, and a 1.6\% uplift of search DAU. We believe the practical insights presented in this work are valuable especially for building and improving personalized search systems for the short video platforms.

\end{abstract}




\maketitle

\section{Introduction} \label{intro}

% P1: breif intro to search engine. 
Search engines serve as an efficient portal for users to promptly locate the information they seek. Traditionally, web search primarily relies on query-document relevance to deliver results, methods for relevance computation include the widely-adopted TF-IDF and BM25\cite{BM25}. Recently, deep learning approaches have gained popularity, especially the pre-trained language models\cite{yates-etal-2021-pretrained, nogueira2019multi,han2020learningtorank,rankT5}. Despite the great performance these models achieved, the query-document retrieve-then-rank paradigm has a pivotal limitation: it overlooks the crucial user context. This neglect may lead to sub-optimal search results, especially when query is ambiguous and user's intentions diverge.

% P2: personalized search. what, why and how.
Personalized search tailor search results to individual needs by incorporating user information beyond the input query. It becomes increasingly popular as search queries in many applications are short and occasionally ambiguous \cite{www07Dou}.Besides, user profiles and historical activities provide valuable information for understanding search intentions and user preferences, ultimately leading to improved search quality. Early attempts to personalized search include constructing and utilizing user profile and past search activities \cite{WSDM10Hassan,WSDM12Sontag,SIGIR12Bennett,CIKM13Harvey,SIGIR14Vu,Vu_2017}. More recently, personalized search extends to a wide range of applications such as social networking \cite{Huang_2020} and e-commerce \cite{zheng2022multiobjective}.

%P3: challenges and opportunities for personalization of content search(e.g., video, img, documents, etc.)
% With the increasing popularity of short video applications, e.g., Tiktok, Kuaishou, Youtube Shorts, etc, it is an interesting and meaningful question to raise that, to what extent does short video search benefit from personalization? We identify some unique opportunities as well as challenges for short video search personalization:

As the popularity of short video applications such as TikTok and Kuaishou continues to soar, a valuable and intriguing question arises: \textbf{can personalization
improve the short-video search engagement, and if so, which techniques
hold the key}? In delving into this research, we uncover two unique opportunities for personalizing short video search:

Firstly, \textbf{the abundance of user watching history}. Short video applications usually have long and abundant user watch histories with various of interests and topics. In our platform, we observe that over 80\% of search users are highly active users of the platform (logging in for more than 20 days per month), and on average they watch over 200 videos each day, most from the recommendation feeds. By leveraging this abundant user watch histories in the platform, we can better understand user long-term interests. Besides, we find more than 1/4 of queries in our platform are issued while users are browsing recommendation feeds, which provides crucial context for understanding user short-term search intent. For example, when a user views a WWDC news conference video and subsequently queries "apple", it is clear the search intention is directed towards the company Apple, not the fruit.

% P5: the short length of queries.
Secondly, \textbf{the brevity of input queries}. We observe that over 40\% of our initiative queries contain less than 6 Chinese characters, which sometimes convey ambiguous search needs and intentions. For example, one of the top search queries "Subject Three" in our platform, can represent both the original meaning, the third subject for driver licence test, or the name for a trendy dancing music. Another instance from our search logs relating to a user searching for "short haircut" and selecting a video tutorial on children's haircuts. While "children haircut" was not explicitly stated in the query "short haircut", the user's past viewing history revealed this implicit need. the brevity of input queries underscores the need for search engines to leverage user context and historical behaviors to disambiguate search intentions.

% P4: Our approach to tackle the issues above. 
In light of these observations, we introduce the \textbf{$\text{PR}^2$} (short for \textbf{P}ersonalized \textbf{R}etrieval and \textbf{R}anking augmented search system), a novel and comprehensive solution for personalizing short video search. In Section 2, we present a brief overview of our personalized search system. Next, aligning with the classic "retrieval-then-rank" IR pipeline, we introduce adaptions of personalized models to both retrieval and ranking. In Section 3, We propose the Query-Relevant Collaborative Filtering (QRCF) and Personalized Dense Retrieval (PDR) methods, which aim at retrieving candidates both relevant to the search query, but also tailored for the user's personal interests. In Section 4, we propose the novel ranking model, namely Query-dominant Interest Network (QIN) , for better utilizing both long-term and short-term user behaviors, and adopt a multi-task learning framework to leverage various user behavior feedback. Then in Section 5, we present the notable A/B testing improvements of deploying the proposed $\text{PR}^2$ solution in Kuaishou, a major short video platform with over 400M DAU. We present the related work in Section 6, and conclude our work in Section 7.

% P5: contribution, 
Overall, our contributions are summarized as follows:
\begin{itemize}
    \item We demonstrate that substantial gains in user engagement can be achieved through personalizing short-video search. We present a compelling case study on Kuaishou, a major short video platform with over 400 million daily active users, showcasing the practical impact and insights gained from deploying our personalized models.
    \item We introduce a comprehensive solution, namely $\text{PR}^2$, for personalizing short-video search. $\text{PR}^2$ seamlessly integrates query-relevant collaborative filtering and personalized dense retrieval, leveraging user behaviors for highly tailored search results. Our QIN ranking model adeptly captures both short-term and long-term user interests, enhanced by a multi-task learning framework that harnesses vast user feedback. To our knowledge, we are the first to propose such a systematic solution for short-video search personalization, offering valuable insights into improving search engagement in real-world applications. 
\end{itemize}

\section{System Overview}

\begin{center}
\begin{figure}[hbtp]
\centering
    \includegraphics[height=8cm]
    {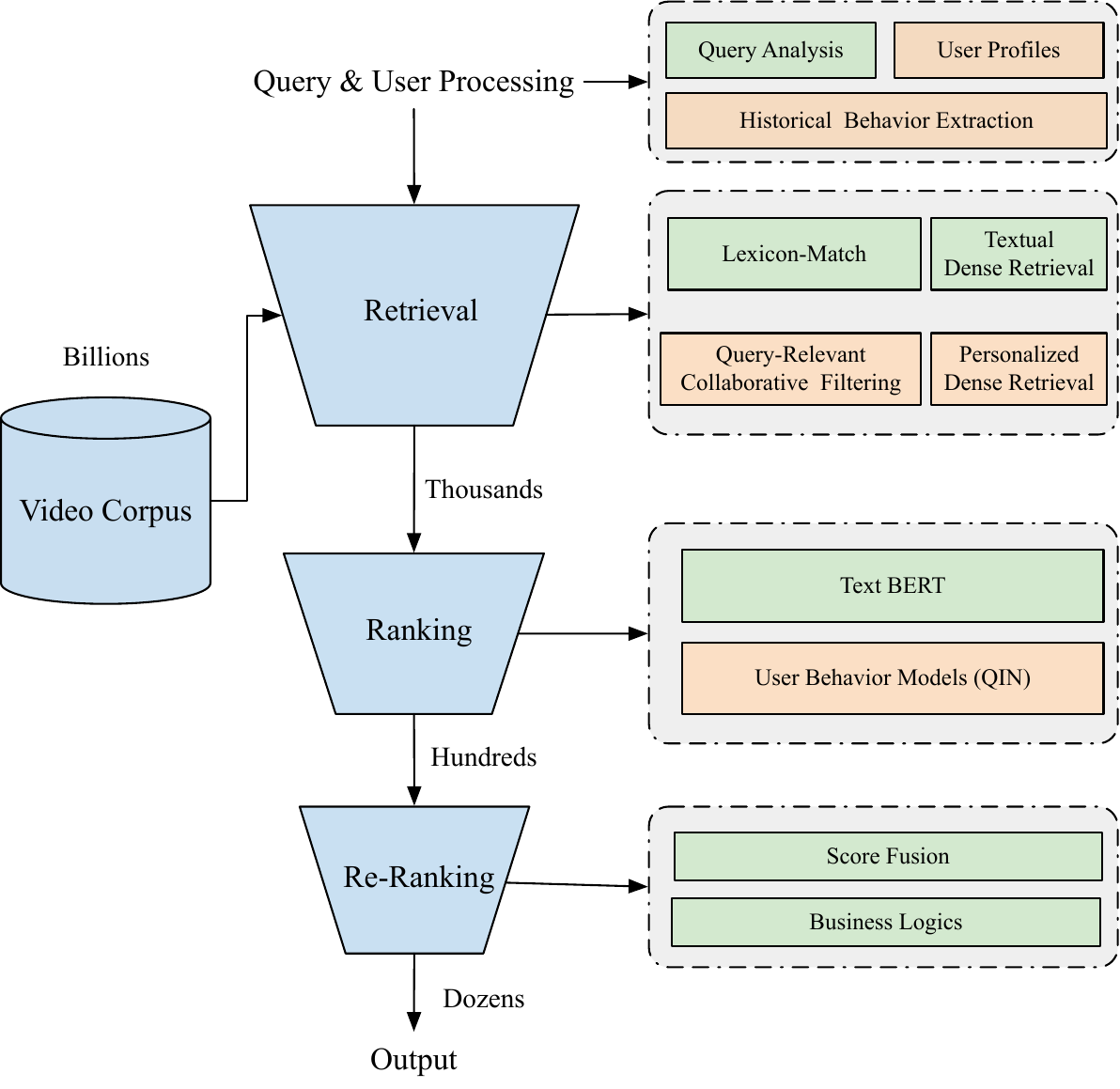}
    \caption{A brief system overview. the traditional "retrieve-then-rank" pipeline is depicted in blue. The existing non-personalized stages are in green, and personalized modules discussed in this work are highlighted in orange.}
    \label{fig:overview}
\end{figure}    
\end{center}

We give a brief architecture overview in Fig. \ref{fig:overview}. To select a dozen satisfactory videos from a billion-scale corpus, our search engine is designed as a multi-stage IR system, which generally comprises three stages: \textit{Retrieval}, \textit{Ranking} and \textit{Re-ranking}.

\textbf{\textit{Retrieval}}. It targets at retrieving thousands of high-quality videos from a billion-scale corpus. Initially, our system relies on content-based retrieval methods, including lexicon match and textual/visual dense retrieval. However, we observe that content-based methods often fall short of retrieving videos with better user engagement. To further improve search experience, we go beyond content-based retrieval and introduce behavior-based methods.

\textbf{\textit{Ranking}}. It targets at generating multiple ranking scores for the thousands of candidates returned from the retrieval. This stage plays a pivotal role in personalized search, as it enables the utilization of richer features and complex model architectures to provide fine-grained ranking. While traditional web search primarily relies on text-based methods like BERT to rank documents, in the context of short-video search, we stress the importance of user behavior ranking models, as it brings significant gain of user engagement and long-term retention.

\textbf{\textit{Re-ranking}}. This stage typically handles candidates in the range of tens to hundreds. Its major task is to fuse multiple ranking scores from the previous stage, and achieve a serious rule-based business logic such as filtering, diversity, etc.

\section{Personalized Retrieval}
\begin{center}
\begin{figure*}[t!]
\centering
\includegraphics[width=0.75\linewidth,height=6.5cm]{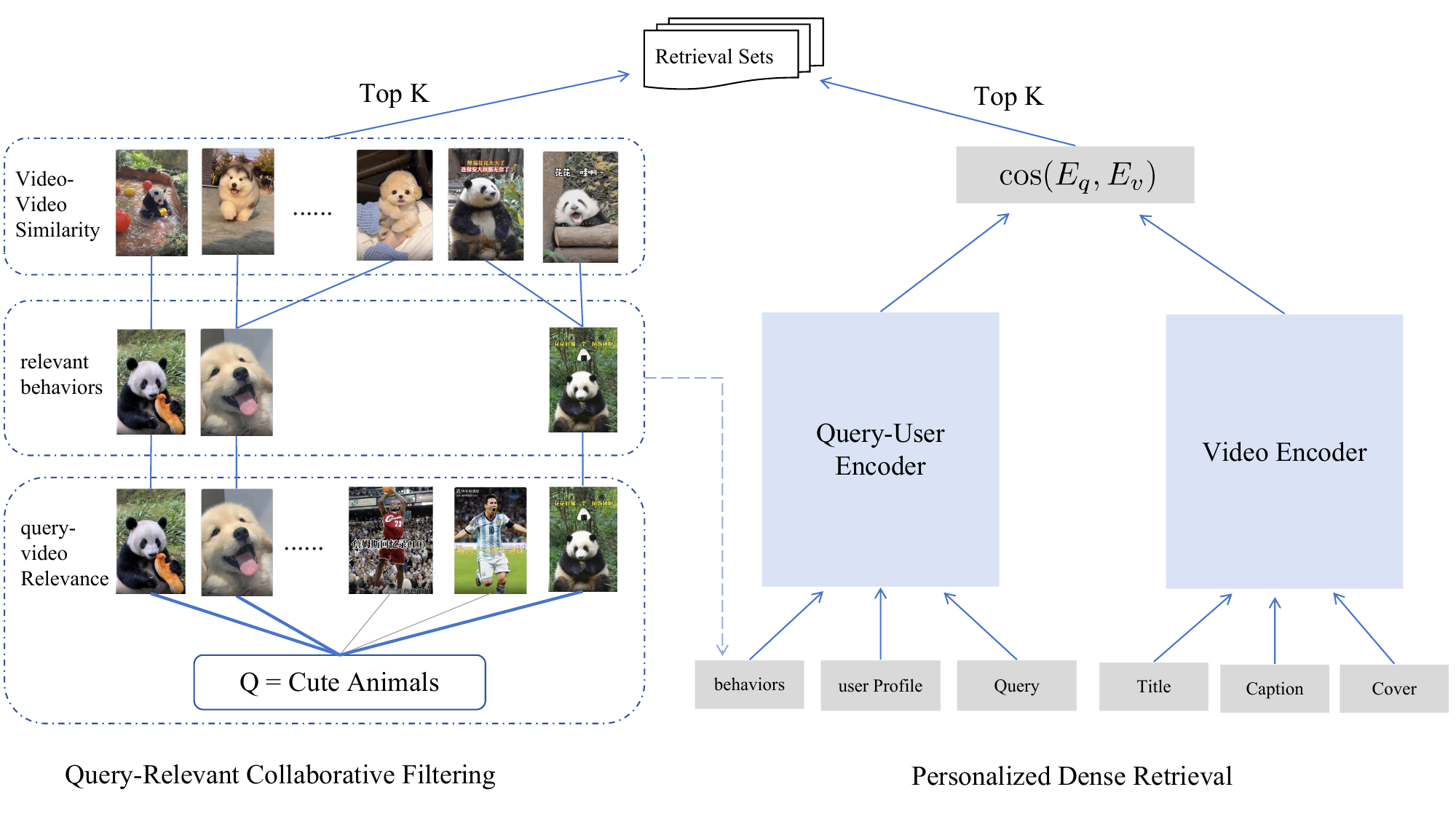}
    \caption{Personalized Retrieval for short video search. We employ collaborative filtering and dense retrieval methods. Each method retrieves topK personalized candidates, then merged and passed to the downstream ranker.}
    \label{fig:search_u2i}
\end{figure*}    
\end{center}

We aim at retrieving videos not only relevant to the issued query, but also based on user profile and historical watching interests. We achieve this goal with two effective methods: Firstly, for the query which users have relevant past watching videos, we leverage these relevant user behaviors, and employ item based collaborative filtering \cite{www01i2i} to retrieve similar candidates. Secondly, for new queries that user do not have relevant behaviors, we encode user and query information into embedding, and leverage dense retrieval methods to generalize and retrieve relevant and personalized candidates. The overall architecture is illustrated in Fig. \ref{fig:search_u2i}.

\subsection{Query-Relevant Collaborative Filtering}
Limited work explores collaborative filtering in search engines, hindered by difficulty in ensuring relevance of retrieved candidates to search queries. To overcome this challenge, We introduce the \textbf{Query-Relevant Collaborative Filtering (QRCF)}. It decomposes the task into two sub-modules: query-video relevance filtering and video-video similarity calculation. 

\subsubsection{Query-Video Relevance Filtering}
Note that users often have engaged with videos across a wide of interests and topics, most of which should not being considered in the current search. To efficiently select the most relevant watching history, we adopt a soft relevance filter. Specifically, based on the previous work \cite{SIGIR22Wang}, we take advantage of a multi-modal embedding model, which comprises of a query encoder $\mathcal{F}_q$ with text as input, and a multi-modal video encoder $\mathcal{F}_v$ with both videos' title and cover as input. It projects queries and videos into the same embedding space, and bridges the representation gap between queries and videos. 

Given the query $q$, user watched video sequence $B = (b_1, b_2, ..., b_T)$, embedding $E_q = \mathcal{F}_q (q)$ and $E_{b_i} = \mathcal{F}_v(b_i)$, we select relevant behaviors,
\begin{equation}
\label{eq:rel_behavior}
    B_{rel} = \{b_i | topk(cos(E_{b_i},E_q), K), cos(E_{b_i},E_q)\geq \epsilon\},
\end{equation}
where $\epsilon$ is a relevance threshold that strikes a balance between the retrieval's relevance and diversity. Higher $\epsilon$ leads to fewer but more relevant behaviors left. K controls how many behaviors we want to utilize. Increasing K yields more candidates but at higher computational cost.

\subsubsection{Video-Video Similarity Calculation}
Various methods can be applied to calculate item similarities \cite{huang2024}. To ensure semantic relevance in the search scenarios, We adopt two classic methods: memory-based and embedding-based item-to-item (I2I). \\
\textbf{Memory-Based I2I.} We use search logs to construct the click graph between users and their clicked videos, and employ the Swing algorithm \cite{yang2020large} to detect robust click co-occurrence among users. To ensure search relevance, we only consider users' click co-occurrence of the same query. The search swing score for item i and j is given by,
\begin{equation}
    s(i,j) = \sum_{u \in S_{i} \cap S_{j}} \sum_{v \in S_{i} \cap S_{j}} \frac{1}{\alpha + |I_{u} \cap I_{v}|},
\end{equation}
where $S_{i}$ denotes the search sessions where users click on item i, $I_{u}$ represents all items clicked in the session u, and $\alpha$ is a smoothing coefficient.\\
\textbf{Embedding-Based I2I}. We also adopt embedding-based collaborative filtering, to calculate similarity of videos that have no co-click in the past search logs. 
% We use the item embedding produced by the dense retrieval model, which will be introduced in Section 3.2, 
Specifically, the embedding-based item-item similarity is,
\begin{equation}
    s(i,j) = cos(E_i, E_j),
\end{equation}
where the item embedding $E_j$ and $E_j$ can come from various video encoders such as the dense retrieval model introduced in Section 3.2. In such case, we firstly store all the video embedding of the dense retrieval model in an ANN server. When serving online, we retrieve top K similar items given each user behavior in $B_{rel}$.

\subsection{Personalized Dense Retrieval}
Dense retrieval is prevalent in search system\cite{liu2021,Huang_2020,zheng2022multiobjective}, to bridge the gap between queries and indexed items. We adapt the dual encoder architecture proposed in \cite{zheng2022multiobjective} to short video search, and focus on personalizing the model by integrating user profile and past behaviors. Besides, we design a multi-objective loss that optimizes for both query-video relevance and user feedback.

\subsubsection{Query-User Encoder}
To personalize the retrieval model, we emphasize the usage of user features in the user-query encoder, including user profiles and past behaviors.\\
\textbf{User profiles.} We select features likely to impact user preferences for video genres and content, including gender, age segment, location, etc. Given the system's limitation in capturing all user details, we further incorporate unique user ID as a sparse feature, and transform it into learnable dense embedding. Notably, user ID embedding significantly outperforms other features, accounting for over 80\% of recall improvements of adding all profile features. We hypothesize that this embedding fine-tunes model's retrieval to each individual user, thereby enhancing the overall personalization.\\
\textbf{User behaviors.} We highlight the usage of user activities, and adopt the attention mechanism to weight different actions. Specifically, from Eq. \ref{eq:rel_behavior} we can get user long-term interests $B_{rel}$. To further mine useful information from the behavior sequence, we use the user profile and query embedding as the \textit{Query} of attention unit, and user activities as the \textit{Value} and \textit{Key}. Then the user behavior representation is given by,
\begin{align}
    E_b &= \text{MultiHeadAtten}(Q,K,V), \\
    Q &= \text{concat}([E_q, E_p])W + b, \\
    K &= V = B_{rel},
\end{align}    
where $E_q$ and $E_p$ denotes the query and user profile embedding, respectively, and the Q, K, V is processed by standard multi-head attention module to produce user behavior embedding $E_b$. \\
\textbf{Query-User Representation}. Given the embedding of user profiles $E_p$, user behaviors $E_b$ and query $E_q$, the personalized representation of query and user is,
\begin{equation}
    E_{qu} = \text{l2\_norm} (\text{MLP}(\text{concat}([E_p, E_b, E_q]))),
\end{equation}
where $\text{MLP}$ represents the standard Multi-Layer Perceptron with the ReLU activation function.

\subsubsection{Multi-Objective Learning.}
To align personalized retrieval with downstream search tasks, we have two major objectives to optimize: \textit{Relevance} and \textit{User Engagement}. As mentioned above, user engagement is measured by positive feedback such as clicks, long views and likes. 
We use sampled softmax to learn each task,
\begin{align}
    L_{\text{o}} &= -\sum_{i=1}^N y_i^{\text{o}} \text{log} \frac{\text{exp}(s_i / \tau)}{\text{exp}(s_i / \tau) + \sum_{j\in N(i)} \text{exp}(s_j / \tau)},
\end{align}
where $s = cos(E_{qu}, E_v)$, $y^{o} \in \{0,1\}$ denotes the binary label, o denotes the objective, N is the mini-batch size, $\tau$ denotes the temperature parameter for softmax loss, and $N(i)$ represents negative samples of i. To enhance model convergence, besides the easy in-batch negatives, we also adopt intra-session hard negatives that pose challenges for both relevance and behavior tasks. Finally, the total loss can be written as,
\begin{equation}
    L_{\text{pdr}} = \sum_{o \in \mathcal{O}} w_o L_{o},
\end{equation}
where $w_o$ as the weight for each training objective, $L_o$ as the training loss, and $o \in \mathcal{O} = \{\text{relevance}, \text{click}, \text{long-play}, \text{like}\}$.

% \begin{table}[]
% \centering
% \caption{User Engagement Gain for Personalized Ranking in the online A/B testing.  * indicates deployed before 2023 and is not include in the Overall results.}
% \label{tab:ranking_gain}
% \resizebox{0.47\textwidth}{!}{
% \begin{tabular}{l|ccc}
% \toprule
% Module & CTR@10 & Watch Time & QCR \\
% \midrule
% Feature engineering&+0.188\%& +0.857\%  & -0.462\% \\ 
% Common RSU* &+2.101\%&+2.936\%&-2.336\% \\
% Target RSU* &+1.078\%&+2.004\%&-1.749\%\\
% CP-RSU &+1.485\%&+0.110\%&-0.851\%\\
% Real-time behaviours &+0.096\%&+2.133\%&-0.889\%\\
% Multi-task learning &+1.798\%&+3.959\%&-1.365\% \\
% \midrule
% Overall in 2023 & +3.12\% & +20\% & -5.47\% \\
% \bottomrule
% \end{tabular}
% }
% \end{table}

\section{Personalized Ranking}
Conventional web search mainly relies on non-personalized scores such as \textit{relevance}, \textit{quality}, \textit{recency} and \textit{authority} to rank results\cite{KDD21Zou,lin2021}. Compared with such non-personalized ranking paradigm, we decompose the ranking of short-video search into two standalone models: \textit{Experience} and \textit{User Engagement}. The experience model produces non-personalized scores such as relevance and quality that of the <q,v> tuples, And the engagement model focuses more on provides personalized ranking scores of <u,q,v> triplets. 

Compared with personalized ranking models in domains like web search and e-commerce, we stress two distinctions in the context of short-video search: \\

\begin{itemize}
    % \item \textit{Fast-Changing Index}: Short video platforms witness an extraordinary influx of tens of millions of fresh videos uploaded daily, the majority of which have a fleeting lifespan. To accurately predict the user feedback of these new videos, the model necessitates not merely utilization of sparse ID features for memorization user preference, but also the integration of content-based embedding features to generalize to unseen data.

    \item \textit{Sparsity of Search Behaviors}: Short video platforms are usually recommendation-centric, i.e., the majority of users engage primarily to consume video recommendations. Consequently, it is pivotal to leverage user activities both in search and recommendations.
    
    \item \textit{Various User Feedback}: Short video platforms have richer types of positive user interactions compared to web search and e-commerce. Thus, personalized models should take advantage of such rich user engagement signals.
    
    % \item \textit{feature engineering}: User behavior models pay more attention to hand-crafted statistics, constructing user profiles, and leveraging user historical behaviors. While relevance models mainly relay on textual features.
    % \item \textit{labels}: User behavior models get positive labels from online search logs, where abundant feedback can be obtained each day. Relevance models are typically pre-trained on web data, and fine-tuned on human labeled in-house dataset.
\end{itemize}

According to the aforementioned distinctions of short video search, we devise the engagement model, named as \textbf{QIN} (Query-dominant Interest Networks), to enhance search personalization. Subsequently, we discuss the overall model architecture, the crafts of leveraging user behaviors, and multi-task learning techniques.

\subsection{Model Architecture}
QIN consists of three building blocks: (1) A feature input layer that transforms numerical and categorical features into learnable embedding, then concatenates them to a single feature representation $E_{\text{f}}^{\text{qin}}$. (2) A behavior modeling module that leverages attention mechanism to process different user behavior sequences, and generates user interests representation $E_{\text{b}}^{\text{qin}}$. (3) An MMOE\cite{KDD18Ma} network that takes feature and user interests representation as input, and optimizes for multiple user feedback labels. In the following, we elaborate the design of user behavior modeling and multi-task learning, the two most pivotal components for personalizing short video search ranking.

\begin{figure*}[ht]
\center
\includegraphics[width=\linewidth]{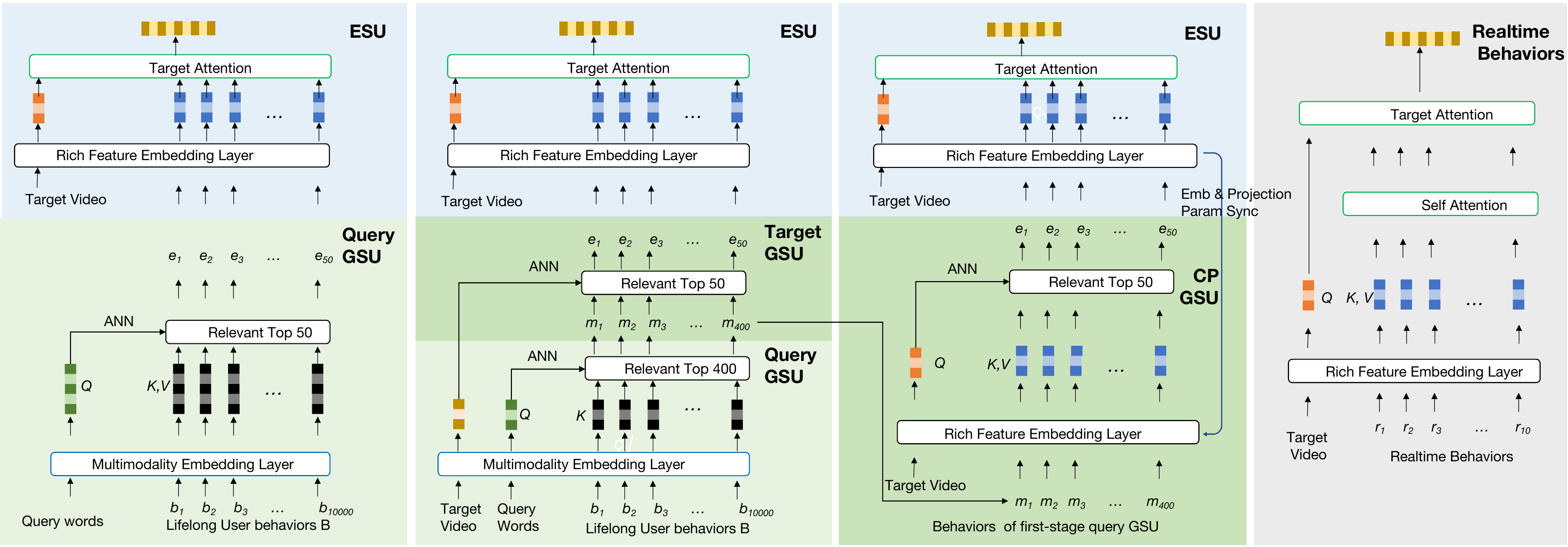}
\caption{Behavior modeling of four user action sequences. From left to right are: query GSU, query-target GSU, CP-GSU, and real-time behaviors. The specific definitions are discussed in Section 4.2.}
\label{fig:seq2}
\end{figure*}

\subsection{Behavior Modeling}
To tackle the sparsity issue, we leverage user behaviors both in search and recommendation feeds. For modeling long-term user interests, we employ the widely-used user interest model SIM\cite{CIKM20Pi}, and tailor the model for search ranking task. For modeling real-time behaviors, we use self attention to extract contextual information, and employ target attention to compare similarity between each behavior with the target video. 

\subsubsection{Long-Term Interests}
SIM decomposes long-term user behavior modeling into two sub-modules: a \textbf{General Search Unit (GSU)} and a \textbf{Exact Search Unit (ESU)}. GSU filters relevant sub-sequence from user long-term behaviors, and ESU calculates fine-grained attention scores on the sub-sequence. We adapt SIM for search ranking by: (1) Leveraging the pre-trained multi-modal encoder, discussed in Section 3.1.1, to enhance GSU's relevance search ability. (2) Designing a two-stage filter for GSU, i.e., filtering firstly by the search query, then by the target video.

Formally, let $B$ denote the life-long user behavior sequence, $q$ the search query, and $t$ the target video. We frame GSU as an ANN task, 
\begin{equation}
 \text{GSU}:=\text{ANN}_E(q,B,K),
\end{equation}
i.e., given query and behavior embedding E, finding the top K behaviors that are most similar to the search query $q$. Based on this formulation, we employ three types of long-term behavior sequences in our model, namely, \textit{Query GSU}, \textit{Query-Target GSU}, \textit{Consistency-Preserved GSU}. The sequences are named after how the GSU is designed.
\begin{itemize}
    \item \textbf{Query GSU}: It filters relevant behaviors with the search query, $B_q = \text{ANN}_{E_1}(q, B, K_1)$.
    \item \textbf{Query-Target GSU}: It adopts a two-stage filter, i.e. filtering firstly by the search query, then by the target video. It makes the sub-sequence relevant to both the search query and target video. $B_{qt} = \text{ANN}_{E_1}(t, \text{ANN}_{E_1}(q,B,K_2), K_1)$.
    \item \textbf{Consistency-Preserved GSU}: It also adopts a two stage query-target filter. But in the second stage of GSU, following the approach of TWIN\cite{KDD23Chang}, it adopts identical embedding for the second-stage of GSU and ESU, making the two modules more compatible. $B_{cp} = \text{ANN}_{E_2}(t, \text{ANN}_{E_1}(q,B,K_2), K_1)$.
\end{itemize}
In practice, K1 and K2 strike a balance between model performance and computational cost. We set K1 = 400 and K2 = 50 in our experiments and find increasing the length yields marginal gain. E1 and E2 denote the embedding of query and videos. We get E1 from the pre-trained multi-modal encoder, and E2 from the ESU. After obtaining the filtered behaviors from GSU, the model then uses the ESU module to calculate fine-grained attention over the target video and relevant behaviors, 
\begin{equation}
 \text{ESU} := \text{TargetAtten}(t, B_{\text{gsu}}),   
\end{equation}
where TargetAtten denotes the standard multi-head attention module with the target video $t$ as the attention query, and selected behaviors $B_{\text{gsu}}$ as the attention key and value. The above long-term behavior modeling modules are illustrated in Figure \ref{fig:seq2}.

% Let $B_q$ denotes the filtered behavior list, then it is retrieved as $B_q = \text{ANN}(q, B, K_1)$. \\

% Let $B_{qt}$ denotes the two-stage filtered behavior list, then it is retrieved as  $B_{qt} = \text{ANN}(t, \text{ANN}(q,B,K_2), K_1)$.\\
% Let $B_{cp}$ denote the filtered behavior list, it is formulated as $B_{cp} = \text{ANN}_{cp}(t, \text{ANN}_{cp}(q,B,K_2), K_1)$, where  $\text{ANN}_{cp}$ simply means the ANN adopts identical embedding from ESU.
% To summarize, the GSU of above three behavior sequences are listed as follows:
% \begin{align}
%     B_q &= \text{ANN}(q, B, K_1), \\
%     B_{qt} &= \text{ANN}(t, \text{ANN}(q,B,K_2), K_1), \\
%     B_{cp} &= \text{ANN}_{cp}(t, \text{ANN}_{cp}(q,B,K_2), K_1),
% \end{align}
% where \text{ANN} adopts pre-trained multi-modal embedding of queries and videos, and $\text{ANN}_{cp}$ adopts embedding from the ESU module.

\subsubsection{Real-Time Behaviors}
Besides the long-term interests, users' real-time actions often convey valuable infromation as well. We utilize the most recent 10 user watching videos, and adapt a self-attention layer to extract contextual information, and employ target attention to compare similarity between each behavior with the target video. The real-time behavior modeling module is illustrated in the most right part in Figure \ref{fig:seq2}.

\subsection{Multi-Task Learning}
We adopt the widely-used MMOE \cite{KDD18Ma} architecture to simultaneously learn multiple user behavior feedback. Formally, suppose QIN learns M tasks, then the output of QIN can be written as,
\begin{equation}
    \mathbf{o}^{\text{qin}} = \text{MMOE}(\text{concat}([E_{\text{rt}}^{\text{qin}},E_{\text{lt}}^{\text{qin}},E_{\text{f}}^{\text{qin}}])),
\end{equation}
where$E_{\text{rt}}^{\text{qin}},E_{\text{lt}}^{\text{qin}},E_{\text{f}}^{\text{qin}}$ denote the long-term user interest embedding, the real-time interest embedding and other categorical and numerical feature embedding. $\mathbf{o}^{\text{qin}}$ denotes the output scores of M tasks. To Train the multi-task model and produce calibrated ranking scores, we adopt the Regression Compatible Ranking (RCR) method, proposed in \cite{bai2023rcr}. The RCR loss of QIN can be written as,
\begin{equation}
    L^{\text{qin}} = \sum_{i=1}^{M} (\text{BCE}(o_i) + \alpha \text{listCE}(o_i)),
\end{equation}
where BCE denotes the binary cross entropy loss, listCE denotes the list-wise cross entropy loss, introduced in \cite{bai2023rcr}, $o_i \in \mathbf{o}^{\text{qin}}$ denotes the calibrated ranking score, and $\alpha$ is the hyper-parameter to balance regression and ranking losses. 

% To fuse all ranking scores, we use the following formula,
% \begin{equation}
%     \text{fused\_rank\_score} = \prod_{i=1}^{M} (1+o_i)^{\alpha_i},
% \end{equation}
% where $o_i \in [0,1]$ belongs to $\mathbf{o}^{\text{qin}}$, and $\alpha_i$ represents the exponents to balance multiple objectives.\\

\subsubsection{Discussion} We find in practice it is valuable to learn more behavior labels, and add them to the ranking formula. The fused ranking score is given by, $\text{fused\_score} = \prod_{i=1}^{M} (1+o_i)^{\alpha_i}$, where $o_i \in \mathbf{o}^{\text{qin}}$. We construct three types of labels from user feedback: \textit{Clicks}, \textit{Play Time}, and \textit{Interactions}. \textit{Play Time} comprises three binary labels (effective play, long play, full play) based on video play time thresholds (7s, 18s, 100\% of video duration). \textit{Interactions} contains binary labels of user explicit feedback such as like, follow, etc. In the experiments, we report online A/B testing of adding the like rate, long-play rate and full-play rate to the ranking formula. All lead to notable engagement improvements.

% In summary, let $\mathbb{Y}$ denote the multi-task label set, we have $\mathbb{Y} = \{y_{clk}, y_{ev}, y_{lv}, y_{pc}, y_{wtq}, y_{like}, y_{follow}, y_{clt}, y_{share}, y_{cmt}\}$.
% To predict the multiple user feedback within a single model, we leverage the widely-adopted MTL framework MMOE \cite{KDD18Ma}. 

% \input{6_balancing_relevance_and_personalization}
\section{Experiments}
In this section, we conduct online experiments to answer to the following research questions:\\
\textbf{RQ1}: Does adding personalized retrieval to the non-personalized system enhance user engagements of short video search?\\
\textbf{RQ2}: Does adding personalized models to search ranking bring significant user engagements of short video search? \\
\textbf{RQ3}: Does the integration of personalized retrieval and ranking modules yield additional advantages over their individual use? \\
% We firstly introduce the metrics we use in practice to access user search experience. Next, we briefly introduce the production baseline model. Then, we move to discuss the experiment results, our key findings, and the ablation study of each module's influence on the overall performance.

\subsection{Evaluation Metrics}
We adopt three types of metrics to comprehensively assess user search satisfaction: \textit{engagement}, \textit{relevance} and \textit{retention}.\\
\textbf{Engagement}: We use three metrics: \textit{CTR@10}, \textit{Video Watch Time per Query (Watch Time)}, and \textit{Like Rate}. 
\begin{itemize}
\item \textit{CTR@10}: This metric represents the click  of the first page (i.e. top 10 positions), defined as $CTR@10 = \frac{\#\text{clicked first page}}{\#\text{search requests}}$. 
% It considers both CTR and the click position, serving as an indicator of whether users can efficiently locate the target result.
\item \textit{Video Watch Time per Query (Watch Time)}: It is defined as $\text{Watch Time per Query} = \frac{\text{total watch time}}{\#\text{query}}$. 
% It accesses whether user's are satisfied with the current search result. 
% \item \textit{Query Change Rate (QCR)}: It is defined as the ratio of change queries to total queries, denoted as $QCR = \frac{\#\text{change query}}{\#\text{query}}$. A change query of the same search intention indicates dissatisfaction with the current search results, thus making the ratio a negative indicator that we strive to minimize.
\item \textit{Like Rate (LR) }: It measures the ratio of likes over all video views. It is defined as $\text{Like Rate} = \frac{\#\text{likes}}{\#\text{video views}}$.
\end{itemize}
\textbf{Relevance}: we introduce the \textit{GSB} \cite{li2023plmRanker} metric to gauge relevance performance. 
\begin{itemize}
\item \textit{Good vs. Same vs. Bad (GSB)}: is a metric that compares two systems in a side-by-side manner. We collect a set of queries from the online search logs and ask expert annotators to give judgments of which system should be more relevant by the users. It is calculated as $\text{GSB} = \frac{\#\text{Good} - \#\text{Bad}}{\#\text{Good} + \#\text{Same} + \#\text{Bad}}$.
\end{itemize}
\textbf{Retention}: Finally, we use \textit{Search Daily Active Users (SDAU)} to gauge long-term user retention. This metric quantifies the search users of our platform. Enhancing this metric is challenging, and it stands as the north star indicator we strive to optimize.

\subsection{Production Base Models}
We introduce the baseline system, a \textbf{non-personalized} retrieve-then-rank search engine.\\
\textbf{Retrieval}: This module consists of two text-based retrieval models, a lexicon-based retriever using BM25\cite{BM25} as the scoring function,  and a two-tower dense retriever.
\begin{itemize}
    \item Lexicon Match (BM25): It retrieves candidates using inverted index, and adopts BM25 as the score function.
    \item Dense Retrieval (DR): It adopts the bi-encoder architecture of ReprBERT \cite{yao22reprBERT}, with distinct 6-layer transformer as query and video encoders. The model only utilizes text features from the query, video title, and caption as input.

\end{itemize}
\textbf{Ranking}: The base ranking model follows the BERT\cite{devlin2019BERT} architecture,  utilizing a 6-layer transformer for fully interactive encoding of query and video textual features. It ranks hundreds of candidates, and each is given a relevance score ranging from 0 to 1.

\subsection{Experiment Settings}
\textbf{Datasets}: The PDR and QIN models undergo weeks of training on vast production search logs, encompassing tens of billions of impressions, billions of clicks, and video views, to achieve convergence. Conversely, the baseline models in production have already achieved full convergence through exhaustive training on historical data.\\
\textbf{Hyper-parameters}: For QRCF, we search the K and $\epsilon$ within the ranges of \{20,50,100\} and \{0.3,0.4,0.5,0.6\}, respectively, and select K=50 and $\epsilon=0.5$ as the optimal values in our system. For each relevant video, we retrieve at most 20 similar candidates, resulting in at most 1000 candidates, from which we select the top 400. For PDR, we adopt three-layer MLPs of dimension $[128,64,32]$ for the query and video encoder, and L2 norm in the top layer. We use cosine similarity as score function, and select the top 100 candidates. For QIN, it employs an MMOE as its backbone, comprising 8 experts, each with dimensions $[512,256,128]$. It encompasses 5 core tasks: click, effective-play, long-play, full-play, like. Each task is assigned a two-layer MLP tower with dimensions $[128,64]$.\\

\subsection{(RQ1) Results of Personalized Retrieval}
Table \ref{tab:pretrieval_results} presents the results of online A/B tests comparing the proposed QRCF and PDR against non-personalized baseline models. 

\begin{center}
\begin{figure}[hbtp]
    \centering
    \includegraphics[width=\linewidth]{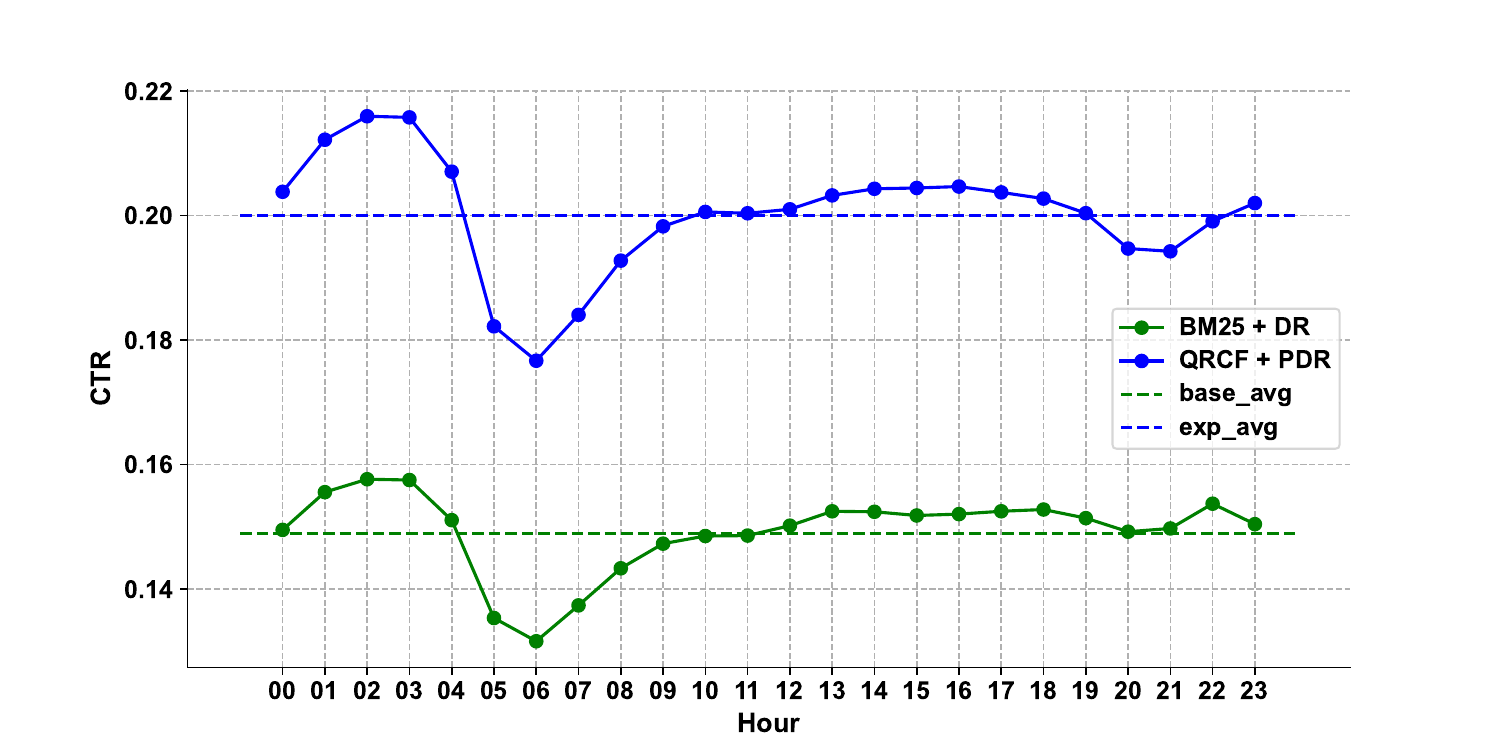}
    \includegraphics[width=\linewidth]
    {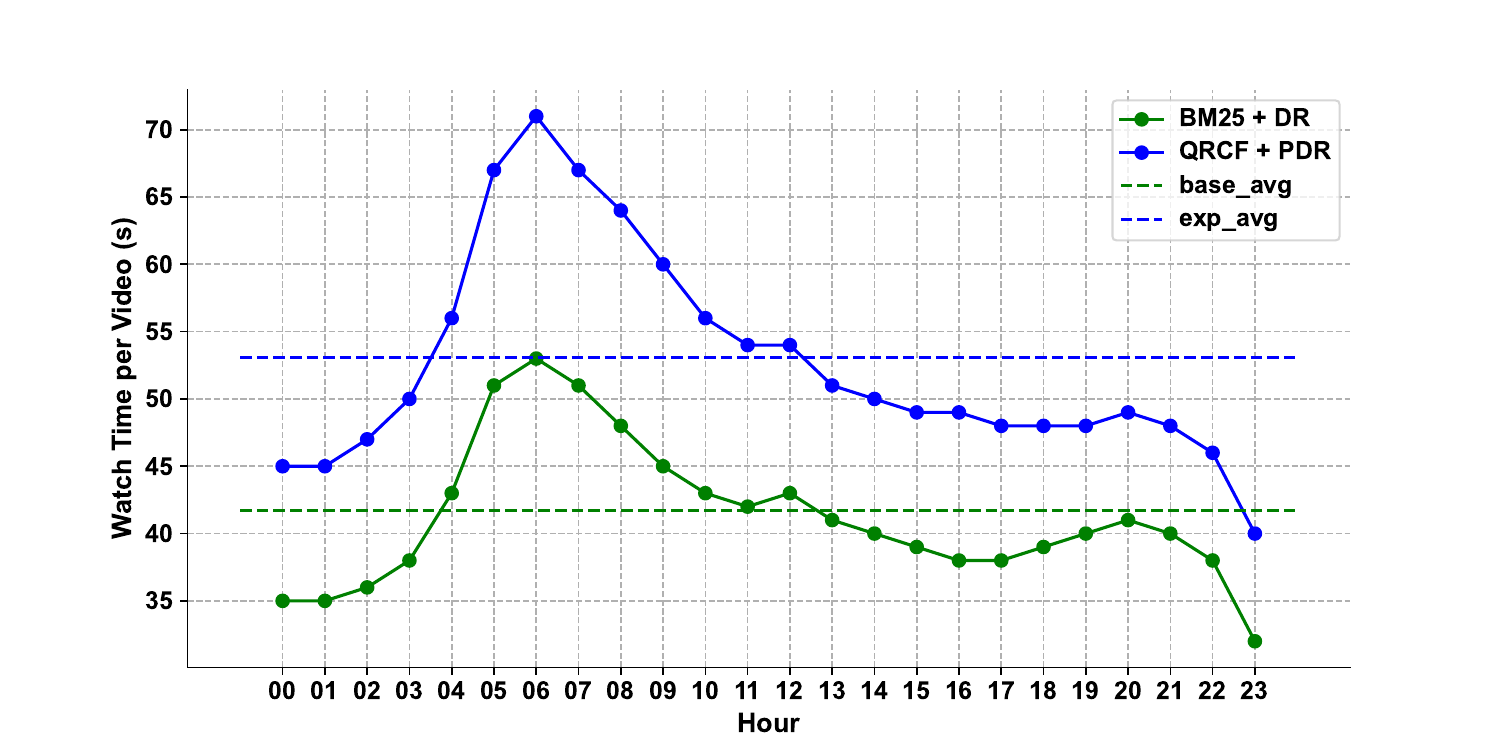}
    \caption{Analysis of online CTR and Watch Time per Video for different retrieval methods. The plot is based on 24-hour online traffic with over 3B video views.}
    \label{fig:pretrieval_ctr_wt}
\end{figure}    
\end{center}

(1) Incorporating personalized retrieval clearly improves user engagement. Overall, We notice a 1.58\% increase in first page CTR, and a 2.39\% increase in video watch time. We owe this improvements to better utilizing user past behaviors, leading to retrieving candidates with better user engagement. To delve deeper, we collect 24 hours' production search logs encompassing over 3B video views, and analyze the empirical CTR and video watch time across different retrieval methods. As shown in Figure \ref{fig:pretrieval_ctr_wt}, we find that personalized retrievers consistently outperform non-personalized methods, averaging a 38\% boost in CTR and 30\% increase in video watch time.

\begin{figure*}[hbtp]
    \centering
    \includegraphics [width=0.8\linewidth]{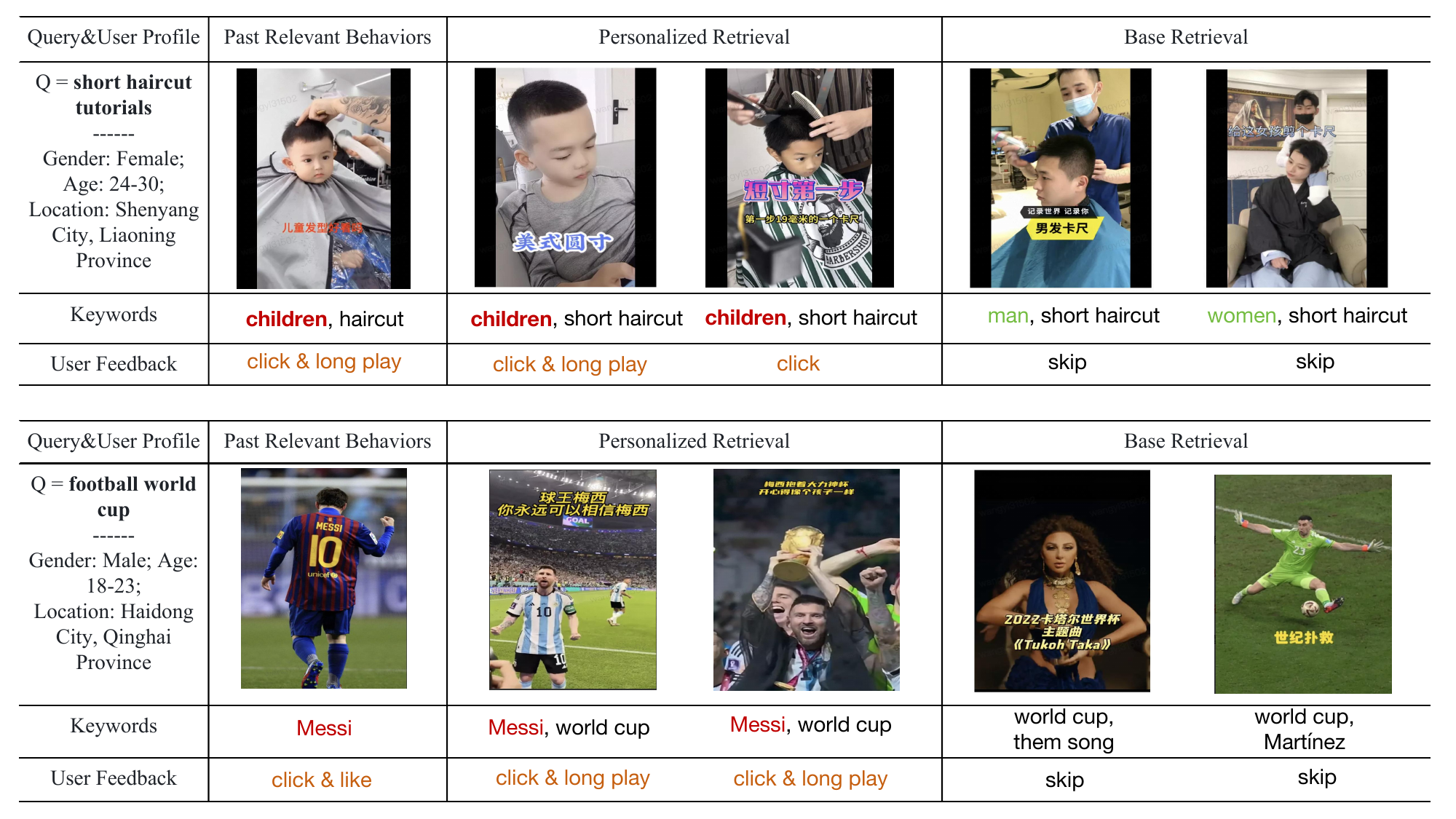}
    \caption{A case study from the online search log. The user's implicit need, highlighted in red, absent from the query but discernible through relevant past behaviors.}
    \label{fig:retrieval_case_study}
\end{figure*}

(2) Incorporating personalized retrieval also improves search relevance. Overall, we notice a 1.9\% increase in GSB, which indicates the search results are more relevant not only to the issued query, but also to the user. We hypothesize the enhanced relevance stems from more accurately capturing users' implicit needs, which are not explicitly stated in queries but discernible from their past behavior. Figure \ref{fig:retrieval_case_study} showcases two good cases we identify from search logs. In the first case, the user searches for "short haircut tutorials". According her past behaviors, we deduce the user's actual intent is seeking tutorials for her children's haircuts. This latent need is effectively captured by personalized retrieval leveraging past actions, yielding relevant results. Conversely, the base model miss such key information. In the second case, the user queries for "football world cup", and engages with videos showcasing Messi's performance on 2022 World Cup. His interest in Messi, evident from his click and like history, is leveraged by personalized retrievals to retrieve relevant results.

\begin{table}[]
\centering
\caption{Weekly online experiments of QRCF and PDR. The base models are non-personalized lexicon-match and embedding-based retrievers. "w/" means integrating the proposed methods into production system.}
\label{tab:pretrieval_results}
\resizebox{0.47\textwidth}{!}{
\begin{tabular}{c|cccc}
\toprule
\multirow{2}{*}{Method}         & \multicolumn{2}{c}{Engagement}      & Relevance       & Retention        \\ \cmidrule(l){2-5} 
                                & CTR@10           & Watch Time       & GSB             & SDAU             \\ \midrule
BM25 + DR                       & -                & -                & -               & -                \\ \midrule
w/ $\text{QRCF}_{\text{swing}}$ & +0.43\%          & +0.84\%          & +0.04\%         & -                \\
w/ $\text{QRCF}_{\text{emb}}$   & +0.62\%          & +0.43\%          & +1.56\%         & -                \\
w/ PDR                          & +0.53\%          & +1.12\%          & +0.3\%          & -                \\ \midrule
w/ QRCF + PDR                         & \textbf{+1.58\%} & \textbf{+2.39\%} & \textbf{+1.9\%} & \textbf{+0.24\%} \\ \bottomrule
\end{tabular}
}
\end{table}

\begin{table*}[hbtp]
\centering
\caption{Weekly online A/B testing for QIN and its key components. The ablation results are presented in the middle rows of the results, whereas the overall performance stands out in bold in the final row. "w/" stands for integrating the proposed module into QIN. The overall performance is tested by integrating QIN into production system.}
\label{tab:ranking_gain}
% \resizebox{0.47\textwidth}{!}{
\begin{tabular}{c|c|cccccc}
\toprule
\multirow{2}{*}{Group} & \multirow{2}{*}{Method} & \multicolumn{4}{c}{Engagement} & Relevance & Retention \\ \cmidrule(l){3-8} 
 &  & CTR@10 & Watch Time & Video Views & Likes & GSB & SDAU \\ \midrule
Production Base & BERT & - & - & - & - & - & - \\ \midrule
\multirow{4}{*}{Behavior Modeling} & w/ Query GSU & {\ul +1.21\%} & +1.58\% & {\ul +1.62\%} & - & +0.5\% & - \\
 & w/ Query-Target GSU & +1.1\% & +1.18\% & +1.28\% & +1.65\% & +2.1\% & - \\
 & w/ CP-RSU & +0.1\% & +0.11\% & +0.87\% & +0.46\% & +0.7\% & - \\
 & w/ Real-Time Behaviors & +0.06\% & +0.46\% & +0.44\% & +0.31\% & +0.7\% & - \\ \midrule
\multirow{3}{*}{Multi-Task Learning} & w/ Like Rate & +0.32\% & +0.94\% & +0.34\% & {\ul +2.06\%} & +0.2\% & - \\
 & w/ Long-Play Rate & +0.31\% & {\ul +1.91\%} & +0.27\% & - & +0.2\% & - \\
 & w/ Full-Play Rate & +0.15\% & +0.23\% & +0.85\% & +0.6\% & -0.2\% & - \\ \midrule
Overall & w/ QIN & \textbf{+4.6\%} & \textbf{+6.5\%} & \textbf{+5.7\%} & \textbf{+5.1\%} & \textbf{+4.2\%} & \textbf{+0.58\%} \\ \bottomrule
\end{tabular}
% }
\end{table*}

\subsection{(RQ2) Results of Personalized Ranking}
Table \ref{tab:ranking_gain} presents the online A/B testing results comparing QIN against baseline models. We offer the following findings:

(1) Incorporating QIN into search ranking brings significant gain of user engagement as well as user retention. Overall, we observe a surge of 4.6\% in CTR@10, 6.5\% in video watch time, and a increase of 5.1\% in video likes. Besides, we also find a notable increase of 0.58\% of search DAU, which serves the north star metric we strive to optimize. Compared with the BERT model which ranks results only based on text features, QIN additionally leverages user profiles, user long-term and short-term behaviors, and statistical features of various timeframe. These abundant ranking features enable QIN better capture user implicit search intent according to behaviors, and predicts more accurate ranking scores.

(2) Comparing various QIN components, we find notable engagement enhancements from behavior modeling and multi-task learning. Notably, common RSU and Target RSU significantly influence CTR@10. We hypothesize that leveraging users' long-term behaviors elevates personalized results, yielding increased clicks and video views. Furthermore, incorporating Long-Play Rate (LPR) and Like-Rate (LR) into ranking models most effectively enhances video watch time and likes metrics. This is attributed to the explicit modeling and utilization of these labels, enabling the search system to prioritize videos with higher LR and LPR, ultimately yielding longer viewing sessions and more likes.

\subsection{(RQ3) Results of Integrating Personalized Retrieval and Ranking}

\begin{table}[]
\centering
\caption{A quarter's online experiments of integrating both personalized retrieval and ranking into production system, to test the long-term effect on user retention.}
\label{tab:abtest}
\resizebox{0.47\textwidth}{!}{
\begin{tabular}{c|cccc} 
\toprule
\multirow{2}{*}{Method} & \multicolumn{2}{c}{Engagement} & Relevance & Retention \\ \cmidrule(l){2-5} 
 & CTR@10 & Watch Time & GSB & SDAU \\ \midrule
Production System & - & - & - & - \\ \midrule
w/ QRCF+PDR & +1.6\% & +2.4\% & +1.9\% & +0.24\% \\
w/ QIN & +4.6\% & +6.5\% & +5.2\% & +0.58\% \\ \midrule
w/ $\text{PR}^2$ & \textbf{+10.2\%} & \textbf{+20\%} & \textbf{+8.1\%} & \textbf{+1.6\%} \\ \bottomrule
\end{tabular}
}
\end{table}

We integrate both proposed personalized retrieval and ranking methods into production system, and summarize the online A/B test results in Table \ref{tab:abtest}. Based on the experiment results, we offer the following findings:

(1) Examining the final row of Table \ref{tab:abtest}, we observe a substantial enhancement in user engagement achieved through personalizing both retrieval and ranking of search system. Specifically, there is a notable 10.2\% increase in the CTR of CTR@10, a substantial 20\% surge in watch time per query, and a 8.1\% increase in GSB. Additionally, we note a promising 1.6\% uplift in search DAU, providing evidence that substantial improvements in user engagement can concurrently bring longer-term user retention.

(2) we observe that the mere summation of metric gains from each stage is less than the figures presented in the final row of Table \ref{tab:abtest}. This suggests that the integration of both personalized retrieval and ranking can yield further metric gains, as the improvements in each stage will mutually reinforce one another.

(3) Comparing retrieval and ranking, we find that personalized ranking yields more prominent improvements, while retrieval attains only 1/3 of the metric gains compared to ranking. We postulate that retrieval, being an upstream component of the system, faces greater challenges in making a direct impact on the final results.

\section{Related Work}
\subsection{Personalized Search}
Personalized search tailors search results to satisfy individual's interest by incorporating user information and past activities. Early studies focusing on how to construct and leverage user profile \cite{SIGIR12Bennett, SIGIR16Cheng, CIKM13Harvey, SIGIR14Vu}. Bennectt et al. \cite{SIGIR12Bennett} assessed how short-term and long-term user behaviors interact, and combine both to improve search quality and personalization. Vu et al. \cite{SIGIR14Vu} proposed a personalisation framework in which a user profile is enriched using information from other users dynamically grouped with respect to an input query. Harvey et al. \cite{CIKM13Harvey} build personalised ranking models in which user profiles are constructed based on the representation of clicked documents over a latent topic space. Cheng et al. \cite{SIGIR16Cheng} proposed novel topic model of constructing latent music interest space, and developed an effective personalized music retrieval system.

More recently, deep learning methods become popular in personalized search, due to its great representation ability, and complex model to fit long-term and dynamic user interests. Great progress has been made in both personalized retrieval and ranking. Here lists a few representative work. Facebook \cite{Huang_2020} applied embedding-based retrieval at social networking search. They introduce unified embedding framework and take into account both query text and searcher's location and social connections. Taobao search personalized their search retrieval and ranking by utilizing user long-term and short-term shopping interactions with context-aware query attention\cite{WSDM22FAN,ni2018perceive,zheng2022multiobjective}. Kuaishou proposed a two-stage query-attention module to filter irrelevant user past behaviors, and improved personalized search ranking\cite{CIKM23Guo}.

The aforementioned studies have made significant progress in enhancing search personalization within individual IR stages, such as user modeling, retrieval, ranking, and re-ranking. However, our work is centered on presenting the successful implementation of a comprehensive, full-stack personalization approach aimed at improving user engagement in the context of short-video search.

\section{Conclusion}
In this work, we comprehensively examine the effort of personalization for a popular short-video platform. We share our experiences adapting retrieval techniques like collaborative filtering and dense retrieval to boost user engagement. We also introduce the behavior model, namely Query-dominant Interest Network (QIN), to accurately predict user feedback. Online A/B tests confirm improved engagement with a 10.2\% CTR@10 increase, and a 20\% surge in video watch time. These insights highlight the significance of personalized search, especially in short video search scenarios.

\bibliographystyle{ACM-Reference-Format}
\bibliography{sample-base}


\begin{thebibliography}{32}

%%% the Web address.

\ifx \showCODEN    \undefined \def \showCODEN     #1{\unskip}     \fi
\ifx \showDOI      \undefined \def \showDOI       #1{#1}\fi
\ifx \showISBNx    \undefined \def \showISBNx     #1{\unskip}     \fi
\ifx \showISBNxiii \undefined \def \showISBNxiii  #1{\unskip}     \fi
\ifx \showISSN     \undefined \def \showISSN      #1{\unskip}     \fi
\ifx \showLCCN     \undefined \def \showLCCN      #1{\unskip}     \fi
\ifx \shownote     \undefined \def \shownote      #1{#1}          \fi
\ifx \showarticletitle \undefined \def \showarticletitle #1{#1}   \fi
\ifx \showURL      \undefined \def \showURL       {\relax}        \fi
% invisible to TeX
\providecommand\bibfield[2]{#2}
\providecommand\bibinfo[2]{#2}
\providecommand\natexlab[1]{#1}
\providecommand\showeprint[2][]{arXiv:#2}

\bibitem[Bai et~al\mbox{.}(2023)]%
        {bai2023rcr}
\bibfield{author}{\bibinfo{person}{Aijun Bai}, \bibinfo{person}{Rolf Jagerman}, \bibinfo{person}{Zhen Qin}, \bibinfo{person}{Le Yan}, \bibinfo{person}{Pratyush Kar}, \bibinfo{person}{Bing-Rong Lin}, \bibinfo{person}{Xuanhui Wang}, \bibinfo{person}{Michael Bendersky}, {and} \bibinfo{person}{Marc Najork}.} \bibinfo{year}{2023}\natexlab{}.
\newblock \bibinfo{title}{Regression Compatible Listwise Objectives for Calibrated Ranking with Binary Relevance}.
\newblock
\newblock
\showeprint[arxiv]{2211.01494}~[cs.IR]
\urldef\tempurl%
\url{https://arxiv.org/abs/2211.01494}
\showURL{%
\tempurl}


\bibitem[Bennett et~al\mbox{.}(2012)]%
        {SIGIR12Bennett}
\bibfield{author}{\bibinfo{person}{Paul~N. Bennett}, \bibinfo{person}{Ryen~W. White}, \bibinfo{person}{Wei Chu}, \bibinfo{person}{Susan~T. Dumais}, \bibinfo{person}{Peter Bailey}, \bibinfo{person}{Fedor Borisyuk}, {and} \bibinfo{person}{Xiaoyuan Cui}.} \bibinfo{year}{2012}\natexlab{}.
\newblock \showarticletitle{Modeling the Impact of Short- and Long-Term Behavior on Search Personalization}. In \bibinfo{booktitle}{\emph{Proceedings of the 35th International ACM SIGIR Conference on Research and Development in Information Retrieval}} (Portland, Oregon, USA) \emph{(\bibinfo{series}{SIGIR '12})}. \bibinfo{publisher}{Association for Computing Machinery}, \bibinfo{address}{New York, NY, USA}, \bibinfo{pages}{185–194}.
\newblock
\showISBNx{9781450314725}
\urldef\tempurl%
\url{https://doi.org/10.1145/2348283.2348312}
\showDOI{\tempurl}


\bibitem[Chang et~al\mbox{.}(2023)]%
        {KDD23Chang}
\bibfield{author}{\bibinfo{person}{Jianxin Chang}, \bibinfo{person}{Chenbin Zhang}, \bibinfo{person}{Zhiyi Fu}, \bibinfo{person}{Xiaoxue Zang}, \bibinfo{person}{Lin Guan}, \bibinfo{person}{Jing Lu}, \bibinfo{person}{Yiqun Hui}, \bibinfo{person}{Dewei Leng}, \bibinfo{person}{Yanan Niu}, \bibinfo{person}{Yang Song}, {and} \bibinfo{person}{Kun Gai}.} \bibinfo{year}{2023}\natexlab{}.
\newblock \showarticletitle{TWIN: TWo-Stage Interest Network for Lifelong User Behavior Modeling in CTR Prediction at Kuaishou}. In \bibinfo{booktitle}{\emph{Proceedings of the 29th ACM SIGKDD Conference on Knowledge Discovery and Data Mining}} (<conf-loc>, <city>Long Beach</city>, <state>CA</state>, <country>USA</country>, </conf-loc>) \emph{(\bibinfo{series}{KDD '23})}. \bibinfo{publisher}{Association for Computing Machinery}, \bibinfo{address}{New York, NY, USA}, \bibinfo{pages}{3785–3794}.
\newblock
\showISBNx{9798400701030}
\urldef\tempurl%
\url{https://doi.org/10.1145/3580305.3599922}
\showDOI{\tempurl}


\bibitem[Cheng et~al\mbox{.}(2016)]%
        {SIGIR16Cheng}
\bibfield{author}{\bibinfo{person}{Zhiyong Cheng}, \bibinfo{person}{Shen Jialie}, {and} \bibinfo{person}{Steven~C.H. Hoi}.} \bibinfo{year}{2016}\natexlab{}.
\newblock \showarticletitle{On Effective Personalized Music Retrieval by Exploring Online User Behaviors} \emph{(\bibinfo{series}{SIGIR '16})}. \bibinfo{publisher}{Association for Computing Machinery}, \bibinfo{address}{New York, NY, USA}, \bibinfo{pages}{125–134}.
\newblock
\showISBNx{9781450340694}
\urldef\tempurl%
\url{https://doi.org/10.1145/2911451.2911491}
\showDOI{\tempurl}


\bibitem[Devlin et~al\mbox{.}(2019)]%
        {devlin2019BERT}
\bibfield{author}{\bibinfo{person}{Jacob Devlin}, \bibinfo{person}{Ming-Wei Chang}, \bibinfo{person}{Kenton Lee}, {and} \bibinfo{person}{Kristina Toutanova}.} \bibinfo{year}{2019}\natexlab{}.
\newblock \bibinfo{title}{BERT: Pre-training of Deep Bidirectional Transformers for Language Understanding}.
\newblock
\newblock
\showeprint[arxiv]{1810.04805}~[cs.CL]
\urldef\tempurl%
\url{https://arxiv.org/abs/1810.04805}
\showURL{%
\tempurl}


\bibitem[Dou et~al\mbox{.}(2007)]%
        {www07Dou}
\bibfield{author}{\bibinfo{person}{Zhicheng Dou}, \bibinfo{person}{Ruihua Song}, {and} \bibinfo{person}{Ji-Rong Wen}.} \bibinfo{year}{2007}\natexlab{}.
\newblock \showarticletitle{A large-scale evaluation and analysis of personalized search strategies}. In \bibinfo{booktitle}{\emph{Proceedings of the 16th International Conference on World Wide Web}} (Banff, Alberta, Canada) \emph{(\bibinfo{series}{WWW '07})}. \bibinfo{publisher}{Association for Computing Machinery}, \bibinfo{address}{New York, NY, USA}, \bibinfo{pages}{581–590}.
\newblock
\showISBNx{9781595936547}
\urldef\tempurl%
\url{https://doi.org/10.1145/1242572.1242651}
\showDOI{\tempurl}


\bibitem[Fan et~al\mbox{.}(2022)]%
        {WSDM22FAN}
\bibfield{author}{\bibinfo{person}{Zhifang Fan}, \bibinfo{person}{Dan Ou}, \bibinfo{person}{Yulong Gu}, \bibinfo{person}{Bairan Fu}, \bibinfo{person}{Xiang Li}, \bibinfo{person}{Wentian Bao}, \bibinfo{person}{Xin-Yu Dai}, \bibinfo{person}{Xiaoyi Zeng}, \bibinfo{person}{Tao Zhuang}, {and} \bibinfo{person}{Qingwen Liu}.} \bibinfo{year}{2022}\natexlab{}.
\newblock \showarticletitle{Modeling Users' Contextualized Page-Wise Feedback for Click-Through Rate Prediction in E-Commerce Search}. In \bibinfo{booktitle}{\emph{Proceedings of the Fifteenth ACM International Conference on Web Search and Data Mining}} (Virtual Event, AZ, USA) \emph{(\bibinfo{series}{WSDM '22})}. \bibinfo{publisher}{Association for Computing Machinery}, \bibinfo{address}{New York, NY, USA}, \bibinfo{pages}{262–270}.
\newblock
\showISBNx{9781450391320}
\urldef\tempurl%
\url{https://doi.org/10.1145/3488560.3498478}
\showDOI{\tempurl}


\bibitem[Guo et~al\mbox{.}(2023)]%
        {CIKM23Guo}
\bibfield{author}{\bibinfo{person}{Tong Guo}, \bibinfo{person}{Xuanping Li}, \bibinfo{person}{Haitao Yang}, \bibinfo{person}{Xiao Liang}, \bibinfo{person}{Yong Yuan}, \bibinfo{person}{Jingyou Hou}, \bibinfo{person}{Bingqing Ke}, \bibinfo{person}{Chao Zhang}, \bibinfo{person}{Junlin He}, \bibinfo{person}{Shunyu Zhang}, \bibinfo{person}{Enyun Yu}, {and} \bibinfo{person}{Wenwu Ou}.} \bibinfo{year}{2023}\natexlab{}.
\newblock \showarticletitle{Query-Dominant User Interest Network for Large-Scale Search Ranking}. In \bibinfo{booktitle}{\emph{Proceedings of the 32nd ACM International Conference on Information and Knowledge Management}} (Birmingham, United Kingdom) \emph{(\bibinfo{series}{CIKM '23})}. \bibinfo{publisher}{Association for Computing Machinery}, \bibinfo{address}{New York, NY, USA}, \bibinfo{pages}{629–638}.
\newblock
\showISBNx{9798400701245}
\urldef\tempurl%
\url{https://doi.org/10.1145/3583780.3615022}
\showDOI{\tempurl}


\bibitem[Han et~al\mbox{.}(2020)]%
        {han2020learningtorank}
\bibfield{author}{\bibinfo{person}{Shuguang Han}, \bibinfo{person}{Xuanhui Wang}, \bibinfo{person}{Mike Bendersky}, {and} \bibinfo{person}{Marc Najork}.} \bibinfo{year}{2020}\natexlab{}.
\newblock \bibinfo{title}{Learning-to-Rank with BERT in TF-Ranking}.
\newblock
\newblock
\showeprint[arxiv]{2004.08476}~[cs.IR]


\bibitem[Harvey et~al\mbox{.}(2013)]%
        {CIKM13Harvey}
\bibfield{author}{\bibinfo{person}{Morgan Harvey}, \bibinfo{person}{Fabio Crestani}, {and} \bibinfo{person}{Mark~J. Carman}.} \bibinfo{year}{2013}\natexlab{}.
\newblock \showarticletitle{Building User Profiles from Topic Models for Personalised Search}. In \bibinfo{booktitle}{\emph{Proceedings of the 22nd ACM International Conference on Information \& Knowledge Management}} (San Francisco, California, USA) \emph{(\bibinfo{series}{CIKM '13})}. \bibinfo{publisher}{Association for Computing Machinery}, \bibinfo{address}{New York, NY, USA}, \bibinfo{pages}{2309–2314}.
\newblock
\showISBNx{9781450322638}
\urldef\tempurl%
\url{https://doi.org/10.1145/2505515.2505642}
\showDOI{\tempurl}


\bibitem[Hassan et~al\mbox{.}(2010)]%
        {WSDM10Hassan}
\bibfield{author}{\bibinfo{person}{Ahmed Hassan}, \bibinfo{person}{Rosie Jones}, {and} \bibinfo{person}{Kristina~Lisa Klinkner}.} \bibinfo{year}{2010}\natexlab{}.
\newblock \showarticletitle{Beyond DCG: User Behavior as a Predictor of a Successful Search}. In \bibinfo{booktitle}{\emph{Proceedings of the Third ACM International Conference on Web Search and Data Mining}} (New York, New York, USA) \emph{(\bibinfo{series}{WSDM '10})}. \bibinfo{publisher}{Association for Computing Machinery}, \bibinfo{address}{New York, NY, USA}, \bibinfo{pages}{221–230}.
\newblock
\showISBNx{9781605588896}
\urldef\tempurl%
\url{https://doi.org/10.1145/1718487.1718515}
\showDOI{\tempurl}


\bibitem[Huang et~al\mbox{.}(2024)]%
        {huang2024}
\bibfield{author}{\bibinfo{person}{Junjie Huang}, \bibinfo{person}{Jizheng Chen}, \bibinfo{person}{Jianghao Lin}, \bibinfo{person}{Jiarui Qin}, \bibinfo{person}{Ziming Feng}, \bibinfo{person}{Weinan Zhang}, {and} \bibinfo{person}{Yong Yu}.} \bibinfo{year}{2024}\natexlab{}.
\newblock \bibinfo{title}{A Comprehensive Survey on Retrieval Methods in Recommender Systems}.
\newblock
\newblock
\showeprint[arxiv]{2407.21022}~[cs.IR]
\urldef\tempurl%
\url{https://arxiv.org/abs/2407.21022}
\showURL{%
\tempurl}


\bibitem[Huang et~al\mbox{.}(2020)]%
        {Huang_2020}
\bibfield{author}{\bibinfo{person}{Jui-Ting Huang}, \bibinfo{person}{Ashish Sharma}, \bibinfo{person}{Shuying Sun}, \bibinfo{person}{Li Xia}, \bibinfo{person}{David Zhang}, \bibinfo{person}{Philip Pronin}, \bibinfo{person}{Janani Padmanabhan}, \bibinfo{person}{Giuseppe Ottaviano}, {and} \bibinfo{person}{Linjun Yang}.} \bibinfo{year}{2020}\natexlab{}.
\newblock \showarticletitle{Embedding-based Retrieval in Facebook Search}. In \bibinfo{booktitle}{\emph{Proceedings of the 26th ACM SIGKDD International Conference on Knowledge Discovery \& Data Mining}} \emph{(\bibinfo{series}{KDD ’20})}. \bibinfo{publisher}{ACM}.
\newblock
\urldef\tempurl%
\url{https://doi.org/10.1145/3394486.3403305}
\showDOI{\tempurl}


\bibitem[Li et~al\mbox{.}(2023)]%
        {li2023plmRanker}
\bibfield{author}{\bibinfo{person}{Canjia Li}, \bibinfo{person}{Xiaoyang Wang}, \bibinfo{person}{Dongdong Li}, \bibinfo{person}{Yiding Liu}, \bibinfo{person}{Yu Lu}, \bibinfo{person}{Shuaiqiang Wang}, \bibinfo{person}{Zhicong Cheng}, \bibinfo{person}{Simiu Gu}, {and} \bibinfo{person}{Dawei Yin}.} \bibinfo{year}{2023}\natexlab{}.
\newblock \bibinfo{title}{Pretrained Language Model based Web Search Ranking: From Relevance to Satisfaction}.
\newblock
\newblock
\showeprint[arxiv]{2306.01599}~[cs.IR]
\urldef\tempurl%
\url{https://arxiv.org/abs/2306.01599}
\showURL{%
\tempurl}


\bibitem[Lin et~al\mbox{.}(2021)]%
        {lin2021}
\bibfield{author}{\bibinfo{person}{Jimmy Lin}, \bibinfo{person}{Rodrigo Nogueira}, {and} \bibinfo{person}{Andrew Yates}.} \bibinfo{year}{2021}\natexlab{}.
\newblock \bibinfo{title}{Pretrained Transformers for Text Ranking: BERT and Beyond}.
\newblock
\newblock
\showeprint[arxiv]{2010.06467}~[cs.IR]
\urldef\tempurl%
\url{https://arxiv.org/abs/2010.06467}
\showURL{%
\tempurl}


\bibitem[Liu et~al\mbox{.}(2021)]%
        {liu2021}
\bibfield{author}{\bibinfo{person}{Yiding Liu}, \bibinfo{person}{Guan Huang}, \bibinfo{person}{Jiaxiang Liu}, \bibinfo{person}{Weixue Lu}, \bibinfo{person}{Suqi Cheng}, \bibinfo{person}{Yukun Li}, \bibinfo{person}{Daiting Shi}, \bibinfo{person}{Shuaiqiang Wang}, \bibinfo{person}{Zhicong Cheng}, {and} \bibinfo{person}{Dawei Yin}.} \bibinfo{year}{2021}\natexlab{}.
\newblock \bibinfo{title}{Pre-trained Language Model for Web-scale Retrieval in Baidu Search}.
\newblock
\newblock
\showeprint[arxiv]{2106.03373}~[cs.IR]
\urldef\tempurl%
\url{https://arxiv.org/abs/2106.03373}
\showURL{%
\tempurl}


\bibitem[Ma et~al\mbox{.}(2018)]%
        {KDD18Ma}
\bibfield{author}{\bibinfo{person}{Jiaqi Ma}, \bibinfo{person}{Zhe Zhao}, \bibinfo{person}{Xinyang Yi}, \bibinfo{person}{Jilin Chen}, \bibinfo{person}{Lichan Hong}, {and} \bibinfo{person}{Ed~H. Chi}.} \bibinfo{year}{2018}\natexlab{}.
\newblock \showarticletitle{Modeling Task Relationships in Multi-task Learning with Multi-gate Mixture-of-Experts}. In \bibinfo{booktitle}{\emph{Proceedings of the 24th ACM SIGKDD International Conference on Knowledge Discovery \& Data Mining}} (London, United Kingdom) \emph{(\bibinfo{series}{KDD '18})}. \bibinfo{publisher}{Association for Computing Machinery}, \bibinfo{address}{New York, NY, USA}, \bibinfo{pages}{1930–1939}.
\newblock
\showISBNx{9781450355520}
\urldef\tempurl%
\url{https://doi.org/10.1145/3219819.3220007}
\showDOI{\tempurl}


\bibitem[Ni et~al\mbox{.}(2018)]%
        {ni2018perceive}
\bibfield{author}{\bibinfo{person}{Yabo Ni}, \bibinfo{person}{Dan Ou}, \bibinfo{person}{Shichen Liu}, \bibinfo{person}{Xiang Li}, \bibinfo{person}{Wenwu Ou}, \bibinfo{person}{Anxiang Zeng}, {and} \bibinfo{person}{Luo Si}.} \bibinfo{year}{2018}\natexlab{}.
\newblock \bibinfo{title}{Perceive Your Users in Depth: Learning Universal User Representations from Multiple E-commerce Tasks}.
\newblock
\newblock
\showeprint[arxiv]{1805.10727}~[stat.ML]


\bibitem[Nogueira et~al\mbox{.}(2019)]%
        {nogueira2019multi}
\bibfield{author}{\bibinfo{person}{Rodrigo Nogueira}, \bibinfo{person}{Wei Yang}, \bibinfo{person}{Kyunghyun Cho}, {and} \bibinfo{person}{Jimmy Lin}.} \bibinfo{year}{2019}\natexlab{}.
\newblock \showarticletitle{Multi-stage document ranking with BERT}.
\newblock \bibinfo{journal}{\emph{arXiv preprint arXiv:1910.14424}} (\bibinfo{year}{2019}).
\newblock


\bibitem[Pi et~al\mbox{.}(2020)]%
        {CIKM20Pi}
\bibfield{author}{\bibinfo{person}{Qi Pi}, \bibinfo{person}{Guorui Zhou}, \bibinfo{person}{Yujing Zhang}, \bibinfo{person}{Zhe Wang}, \bibinfo{person}{Lejian Ren}, \bibinfo{person}{Ying Fan}, \bibinfo{person}{Xiaoqiang Zhu}, {and} \bibinfo{person}{Kun Gai}.} \bibinfo{year}{2020}\natexlab{}.
\newblock \showarticletitle{Search-Based User Interest Modeling with Lifelong Sequential Behavior Data for Click-Through Rate Prediction}. In \bibinfo{booktitle}{\emph{Proceedings of the 29th ACM International Conference on Information \& Knowledge Management}} (Virtual Event, Ireland) \emph{(\bibinfo{series}{CIKM '20})}. \bibinfo{publisher}{Association for Computing Machinery}, \bibinfo{address}{New York, NY, USA}, \bibinfo{pages}{2685–2692}.
\newblock
\showISBNx{9781450368599}
\urldef\tempurl%
\url{https://doi.org/10.1145/3340531.3412744}
\showDOI{\tempurl}


\bibitem[Robertson and Zaragoza(2009)]%
        {BM25}
\bibfield{author}{\bibinfo{person}{Stephen Robertson} {and} \bibinfo{person}{Hugo Zaragoza}.} \bibinfo{year}{2009}\natexlab{}.
\newblock \showarticletitle{The Probabilistic Relevance Framework: BM25 and Beyond}.
\newblock \bibinfo{journal}{\emph{Found. Trends Inf. Retr.}} \bibinfo{volume}{3}, \bibinfo{number}{4} (\bibinfo{date}{apr} \bibinfo{year}{2009}), \bibinfo{pages}{333–389}.
\newblock
\showISSN{1554-0669}
\urldef\tempurl%
\url{https://doi.org/10.1561/1500000019}
\showDOI{\tempurl}


\bibitem[Sarwar et~al\mbox{.}(2001)]%
        {www01i2i}
\bibfield{author}{\bibinfo{person}{Badrul Sarwar}, \bibinfo{person}{George Karypis}, \bibinfo{person}{Joseph Konstan}, {and} \bibinfo{person}{John Riedl}.} \bibinfo{year}{2001}\natexlab{}.
\newblock \showarticletitle{Item-based collaborative filtering recommendation algorithms}. In \bibinfo{booktitle}{\emph{Proceedings of the 10th International Conference on World Wide Web}} (Hong Kong, Hong Kong) \emph{(\bibinfo{series}{WWW '01})}. \bibinfo{publisher}{Association for Computing Machinery}, \bibinfo{address}{New York, NY, USA}, \bibinfo{pages}{285–295}.
\newblock
\showISBNx{1581133480}
\urldef\tempurl%
\url{https://doi.org/10.1145/371920.372071}
\showDOI{\tempurl}


\bibitem[Sontag et~al\mbox{.}(2012)]%
        {WSDM12Sontag}
\bibfield{author}{\bibinfo{person}{David Sontag}, \bibinfo{person}{Kevyn Collins-Thompson}, \bibinfo{person}{Paul~N. Bennett}, \bibinfo{person}{Ryen~W. White}, \bibinfo{person}{Susan Dumais}, {and} \bibinfo{person}{Bodo Billerbeck}.} \bibinfo{year}{2012}\natexlab{}.
\newblock \showarticletitle{Probabilistic Models for Personalizing Web Search}. In \bibinfo{booktitle}{\emph{Proceedings of the Fifth ACM International Conference on Web Search and Data Mining}} (Seattle, Washington, USA) \emph{(\bibinfo{series}{WSDM '12})}. \bibinfo{publisher}{Association for Computing Machinery}, \bibinfo{address}{New York, NY, USA}, \bibinfo{pages}{433–442}.
\newblock
\showISBNx{9781450307475}
\urldef\tempurl%
\url{https://doi.org/10.1145/2124295.2124348}
\showDOI{\tempurl}


\bibitem[Vu et~al\mbox{.}(2017)]%
        {Vu_2017}
\bibfield{author}{\bibinfo{person}{Thanh Vu}, \bibinfo{person}{Dat~Quoc Nguyen}, \bibinfo{person}{Mark Johnson}, \bibinfo{person}{Dawei Song}, {and} \bibinfo{person}{Alistair Willis}.} \bibinfo{year}{2017}\natexlab{}.
\newblock \bibinfo{booktitle}{\emph{Search Personalization with Embeddings}}.
\newblock \bibinfo{publisher}{Springer International Publishing}, \bibinfo{pages}{598–604}.
\newblock
\showISBNx{9783319566085}
\showISSN{1611-3349}
\urldef\tempurl%
\url{https://doi.org/10.1007/978-3-319-56608-5_54}
\showDOI{\tempurl}


\bibitem[Vu et~al\mbox{.}(2014)]%
        {SIGIR14Vu}
\bibfield{author}{\bibinfo{person}{Thanh~Tien Vu}, \bibinfo{person}{Dawei Song}, \bibinfo{person}{Alistair Willis}, \bibinfo{person}{Son~Ngoc Tran}, {and} \bibinfo{person}{Jingfei Li}.} \bibinfo{year}{2014}\natexlab{}.
\newblock \showarticletitle{Improving Search Personalisation with Dynamic Group Formation}. In \bibinfo{booktitle}{\emph{Proceedings of the 37th International ACM SIGIR Conference on Research \& Development in Information Retrieval}} (Gold Coast, Queensland, Australia) \emph{(\bibinfo{series}{SIGIR '14})}. \bibinfo{publisher}{Association for Computing Machinery}, \bibinfo{address}{New York, NY, USA}, \bibinfo{pages}{951–954}.
\newblock
\showISBNx{9781450322577}
\urldef\tempurl%
\url{https://doi.org/10.1145/2600428.2609482}
\showDOI{\tempurl}


\bibitem[Wang et~al\mbox{.}(2022)]%
        {SIGIR22Wang}
\bibfield{author}{\bibinfo{person}{Xun Wang}, \bibinfo{person}{Bingqing Ke}, \bibinfo{person}{Xuanping Li}, \bibinfo{person}{Fangyu Liu}, \bibinfo{person}{Mingyu Zhang}, \bibinfo{person}{Xiao Liang}, {and} \bibinfo{person}{Qiushi Xiao}.} \bibinfo{year}{2022}\natexlab{}.
\newblock \showarticletitle{Modality-Balanced Embedding for Video Retrieval}. In \bibinfo{booktitle}{\emph{Proceedings of the 45th International ACM SIGIR Conference on Research and Development in Information Retrieval}} (<conf-loc>, <city>Madrid</city>, <country>Spain</country>, </conf-loc>) \emph{(\bibinfo{series}{SIGIR '22})}. \bibinfo{publisher}{Association for Computing Machinery}, \bibinfo{address}{New York, NY, USA}, \bibinfo{pages}{2578–2582}.
\newblock
\showISBNx{9781450387323}
\urldef\tempurl%
\url{https://doi.org/10.1145/3477495.3531899}
\showDOI{\tempurl}


\bibitem[Yang et~al\mbox{.}(2020)]%
        {yang2020large}
\bibfield{author}{\bibinfo{person}{Xiaoyong Yang}, \bibinfo{person}{Yadong Zhu}, \bibinfo{person}{Yi Zhang}, \bibinfo{person}{Xiaobo Wang}, {and} \bibinfo{person}{Quan Yuan}.} \bibinfo{year}{2020}\natexlab{}.
\newblock \bibinfo{title}{Large Scale Product Graph Construction for Recommendation in E-commerce}.
\newblock
\newblock
\showeprint[arxiv]{2010.05525}~[cs.IR]


\bibitem[Yao et~al\mbox{.}(2022)]%
        {yao22reprBERT}
\bibfield{author}{\bibinfo{person}{Shaowei Yao}, \bibinfo{person}{Jiwei Tan}, \bibinfo{person}{Xi Chen}, \bibinfo{person}{Juhao Zhang}, \bibinfo{person}{Xiaoyi Zeng}, {and} \bibinfo{person}{Keping Yang}.} \bibinfo{year}{2022}\natexlab{}.
\newblock \showarticletitle{ReprBERT: Distilling BERT to an Efficient Representation-Based Relevance Model for E-Commerce}. In \bibinfo{booktitle}{\emph{Proceedings of the 28th ACM SIGKDD Conference on Knowledge Discovery and Data Mining}} (Washington DC, USA) \emph{(\bibinfo{series}{KDD '22})}. \bibinfo{publisher}{Association for Computing Machinery}, \bibinfo{address}{New York, NY, USA}, \bibinfo{pages}{4363–4371}.
\newblock
\showISBNx{9781450393850}
\urldef\tempurl%
\url{https://doi.org/10.1145/3534678.3539090}
\showDOI{\tempurl}


\bibitem[Yates et~al\mbox{.}(2021)]%
        {yates-etal-2021-pretrained}
\bibfield{author}{\bibinfo{person}{Andrew Yates}, \bibinfo{person}{Rodrigo Nogueira}, {and} \bibinfo{person}{Jimmy Lin}.} \bibinfo{year}{2021}\natexlab{}.
\newblock \showarticletitle{Pretrained Transformers for Text Ranking: {BERT} and Beyond}. In \bibinfo{booktitle}{\emph{Proceedings of the 2021 Conference of the North American Chapter of the Association for Computational Linguistics: Human Language Technologies: Tutorials}}, \bibfield{editor}{\bibinfo{person}{Greg Kondrak}, \bibinfo{person}{Kalina Bontcheva}, {and} \bibinfo{person}{Dan Gillick}} (Eds.). \bibinfo{publisher}{Association for Computational Linguistics}, \bibinfo{address}{Online}, \bibinfo{pages}{1--4}.
\newblock
\urldef\tempurl%
\url{https://doi.org/10.18653/v1/2021.naacl-tutorials.1}
\showDOI{\tempurl}


\bibitem[Zheng et~al\mbox{.}(2022)]%
        {zheng2022multiobjective}
\bibfield{author}{\bibinfo{person}{Yukun Zheng}, \bibinfo{person}{Jiang Bian}, \bibinfo{person}{Guanghao Meng}, \bibinfo{person}{Chao Zhang}, \bibinfo{person}{Honggang Wang}, \bibinfo{person}{Zhixuan Zhang}, \bibinfo{person}{Sen Li}, \bibinfo{person}{Tao Zhuang}, \bibinfo{person}{Qingwen Liu}, {and} \bibinfo{person}{Xiaoyi Zeng}.} \bibinfo{year}{2022}\natexlab{}.
\newblock \bibinfo{title}{Multi-Objective Personalized Product Retrieval in Taobao Search}.
\newblock
\newblock
\showeprint[arxiv]{2210.04170}~[cs.IR]


\bibitem[Zhuang et~al\mbox{.}(2023)]%
        {rankT5}
\bibfield{author}{\bibinfo{person}{Honglei Zhuang}, \bibinfo{person}{Zhen Qin}, \bibinfo{person}{Rolf Jagerman}, \bibinfo{person}{Kai Hui}, \bibinfo{person}{Ji Ma}, \bibinfo{person}{Jing Lu}, \bibinfo{person}{Jianmo Ni}, \bibinfo{person}{Xuanhui Wang}, {and} \bibinfo{person}{Michael Bendersky}.} \bibinfo{year}{2023}\natexlab{}.
\newblock \showarticletitle{RankT5: Fine-Tuning T5 for Text Ranking with Ranking Losses}. In \bibinfo{booktitle}{\emph{Proceedings of the 46th International ACM SIGIR Conference on Research and Development in Information Retrieval}} \emph{(\bibinfo{series}{SIGIR '23})}. \bibinfo{publisher}{Association for Computing Machinery}, \bibinfo{address}{New York, NY, USA}, \bibinfo{pages}{2308–2313}.
\newblock
\showISBNx{9781450394086}
\urldef\tempurl%
\url{https://doi.org/10.1145/3539618.3592047}
\showDOI{\tempurl}


\bibitem[Zou et~al\mbox{.}(2021)]%
        {KDD21Zou}
\bibfield{author}{\bibinfo{person}{Lixin Zou}, \bibinfo{person}{Shengqiang Zhang}, \bibinfo{person}{Hengyi Cai}, \bibinfo{person}{Dehong Ma}, \bibinfo{person}{Suqi Cheng}, \bibinfo{person}{Shuaiqiang Wang}, \bibinfo{person}{Daiting Shi}, \bibinfo{person}{Zhicong Cheng}, {and} \bibinfo{person}{Dawei Yin}.} \bibinfo{year}{2021}\natexlab{}.
\newblock \showarticletitle{Pre-trained Language Model based Ranking in Baidu Search}. In \bibinfo{booktitle}{\emph{Proceedings of the 27th ACM SIGKDD Conference on Knowledge Discovery \& Data Mining}} (Virtual Event, Singapore) \emph{(\bibinfo{series}{KDD '21})}. \bibinfo{publisher}{Association for Computing Machinery}, \bibinfo{address}{New York, NY, USA}, \bibinfo{pages}{4014–4022}.
\newblock
\showISBNx{9781450383325}
\urldef\tempurl%
\url{https://doi.org/10.1145/3447548.3467147}
\showDOI{\tempurl}


\end{thebibliography}

\appendix

\end{document}